# Thermoelectric performance of P-N-P abrupt heterostructures vertical to temperature gradient

Bohang Nan, Guiying Xu*, Quanxin Yang and Tao Guo

Beijing Municipal Key Lab of Advanced Energy Materials and Technology, School of Materials Science and Engineering, University of Science and Technology Beijing, Beijing 100083, China

Corresponding authors: xugy@mater.ustb.edu.cn (G. Xu)

Abstract.

We present a model for P-N-P abrupt heterostructures vertical to temperature gradient to improve the thermoelectric performance. The P-N-P heterostructure is considered as an abrupt bipolar junction transistor due to an externally applied temperature gradient paralleled to depletion layers. Taking $Bi_2Te_3$ and $Bi_{0.5}Sb_{1.5}Te_3$ as N-type and P-type thermoelectric materials respectively for example, we achieve the purpose of controlling the Seebeck coefficient and the electrical conductivity independently while amplifying operation power. The calculated results show that the Seebeck coefficient can reach 3312μV/K, and the $ZT_{max}$ values of this model are 45 or 425, which are tens or even hundreds of times greater than those of bulk materials and films.

# I. Introduction

It is well known that energy issue is imperative for most countries in the world. Many researchers have searched for novel sustainable and renewable energy sources in order to alleviate energy consumption needs and increase utilization efficiency. Thermoelectric materials have attracted a lot of attention because of their reversible conversion between heat and electricity.

Thermoelectric performance is quantified by the dimensionless figure of merit $ZT = S^2\sigma T/\kappa$, where $S$ is the Seebeck coefficient, $\sigma$ is the electrical conductivity, $T$ is the absolute temperature and $\kappa$ is the thermal conductivity.[1] Thermal conductivity primarily includes the lattice thermal conductivity ($\kappa_L$) and the electronic thermal conductivity ($\kappa_e$). , $S^2\sigma$ is called "power factor" of the thermoelectric material.

However, these parameters are mutually influential, which makes it difficult to enhance thermoelectric performance. Therefore, most traditional bulk thermoelectric materials can reach $ZT_{\max} \approx 1$, such as $Bi_2Te_3$, PbTe and SiGe alloys.[2]

Generally speaking, if the commercial thermoelectric materials have a higher ZT, their conversion efficiency would be higher.[3] And thermoelectric materials need have ZT values greater than 3 to be competitive with traditional mechanical equipment.[4]

In order to enhance ZT values, low-dimensional thermoelectric materials have become the forefront of research since 1990s. Hicks and Dresselhaus initially presented the concept that low-dimensional materials could increase their thermoelectric performance after improving its power factor in 1993.[5] Based on their assumptions, a $ZT_{\max} = 6$ for the $Bi_2Te_3$ two-dimensional material was obtained. This prediction was confirmed by many experimental data. The high $ZT_{\max}$ value of 2.4 in P-type $Bi_2Te_3/Sb_2Te_3$ quantum-well superlattice material was prepared by Venkatasubramanian et al.[6] Then a greatly enhanced $ZT_{\max}$ value of 3 was obtained in N-type PbSeTe/PbTe quantum-dot superlattice material by Harman et al.[7] In addition, $ZT = 1$ silicon nanowires achieved by Boukai et al were one hundred times greater than bulk silicon materials.[8] Hinterleitner et al. prepared $Fe_2V_{0.8}W_{0.2}Al$ films by magnetron sputtering, which reached a maximum $ZT_{\max}$ of 6 around 350-400K.[9]

These enhanced thermoelectric performances mainly originated from two aspects: (1) the Seebeck coefficient can be improved by introducing quantum confinement effect;[10,11] (2) the lattice thermal conductivity can be decreased because phonons are effectively scattered by interfaces.[12-14] However, there are some questions that hinder the development of low-dimensional thermoelectric materials. First of all, the improved $ZT$ values of most low-dimensional thermoelectric materials are derived from the reduction of the lattice thermal conductivity, rather than the increase of power factor. Secondly, it is difficult to control S and σ independently. Therefore, it is a more common choice to make the addition of the Seebeck coefficient greater than the reduction of the electrical conductivity.

In order to solve these questions, we present a model for the P-N-P heterostructure, which was designed through arranging several P-N junction thermoelectric materials in a certain structure. In fact, P-N junctions have been studied a lot and used widely since 1950s. In the late 1990s, P-N junctions were predicted to enhance Seebeck coefficients and $ZT$ values of thermoelectric materials by Dashevsky,[15] Zakhidov,[16] Ravich[17] and others. Wagner and Span proposed a new approach that P-type material and N-type material could be directly connected without metal electrodes. Meanwhile, a temperature gradient was applied along this P-N junction, as shown in Figure 1.[18] The measured results demonstrated that the efficiency of this approach is higher than traditional P-N junctions.[19] Further research on this approach was carried out by Fu Deyi et al. Fu Deyi theorized that this approach could increase the Seebeck coefficient and decrease the electronic thermal conductivity due to the effect of eddy currents in the P-N junction.[20]

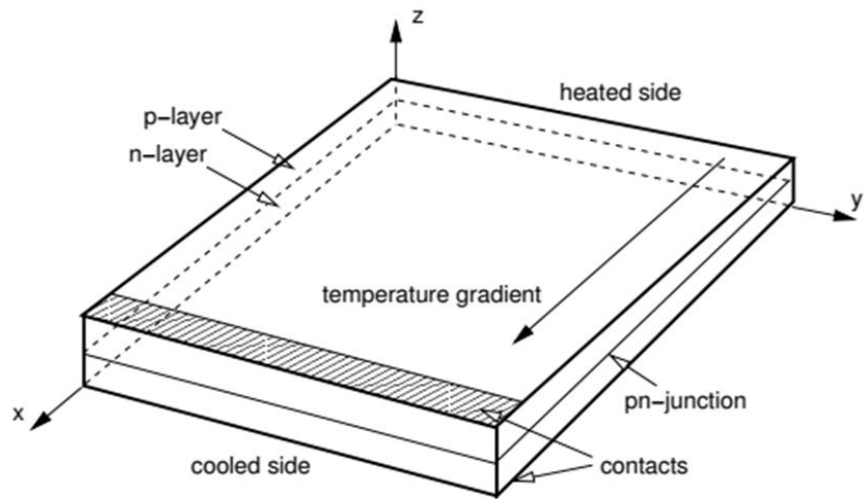

Figure 1. P-N junction with applied temperature gradient

In this work, we propose a P-N-P abrupt heterostructure model vertical to temperature gradient, and provide the detailed analysis process along with calculated results. Here, due to the effect of temperature gradient, the P-N-P heterostructure behaves as a bipolar junction transistor to achieve power amplification. The collector current is controlled by the emitter current in a bipolar transistor, while the Seebeck coefficient is controlled by base-collector interface. The purpose of independent control with regards to both the Seebeck coefficient and electrical conductivity is realized. As a result, the $ZT_{max}$ values can reach tens or even hundreds of times greater than those of bulk materials and films.

The paper is organized as follow: In Sec. II we shall describe our model for P-N-P heterostrctures. In Sec. III we will show and discuss our calculated results. In Sec. IV we shall conclude.

The scheme that has been adopted with respect to notation is illustrated by the partial list of symbols shown in Table I. In addition, the symbol μ has been employed for charge carrier mobility, D for diffusion constant.

Table I. Partial list of symbols used

| Symbol | meaning |
|---|---|
| $N_a$, $N_d$ | Doping concentration in a P-type material, in a N-type material. |
| $E_x$ | Electric field intensity in an abrupt P-N junction. |
| $x_p$, $x_n$ | Depletion layer width near a P-type region, near a N-type region. |
| φ | Electric potential variation in an abrupt P-N junction |
| $\varepsilon_r$ | Relative permittivity of the material |
| $\varepsilon_0$ | Vacuum dielectric constant |
| q | the electronic charge, $=1.6\times10^{-19}$C |
| $k_B$ | the Boltzmann constant, $=1.38\times10^{-23}$J/K |
| $n_i$ | Intrinsic carrier concentration of the material |
| $V_s$ | The Seebeck voltage generated by the temperature gradient. |
| $E_C$, $E_V$ | Conduction band minimum of the material, valence band maximum of the material. |
| $E_{Fp}$, $E_{Fn}$ | The quasi-Fermi level in a P-type material, in a N-type material. |
| m* | The effective mass of the material. |

# II. Analytical model for P-N-P heterostructures

An analytical model for thermoelectric properties of P-N-P heterostructure is presented in this section. Under the effect of temperature gradient, P-N-P heterostructures behave as a bipolar transistor. A general circuit diagram of a P-N-P heterostructure is shown in Figure 2, where P-type material (at the left side of this figure) is considered as an emitter material, N-type material is treated as a base material, and another P-type material (at the right side of this figure) is acts as a collector material. In addition, there are depletion layers at the both of base-emitter interface and at the base-collector interface. The base width is considered to be the sum of the width of both depletion layers near N-type region. As a result, the base region can serve as an insulating layer, in which majority carriers have been depleted. The direction of temperature gradient is assumed to be parallel to these depletion layers. Therefore, in the X direction, the Seebeck voltage can be produced due to the existence of temperature gradient. Meanwhile, the Seebeck voltage can become an external power supply in the bipolar transistor. We have biased the emitter forward and the collector in reverse to achieve the purpose of power amplification. We have also treated the bipolar transistor as two coupled P-N diodes, according to the Ebers-Moll model.[21]

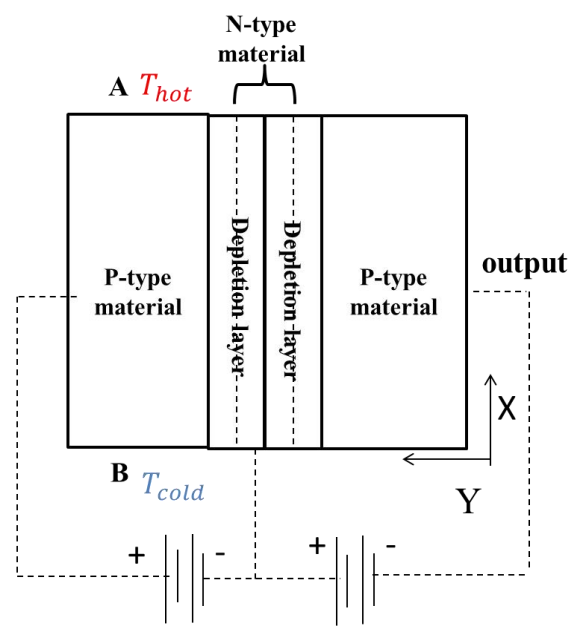

Figure 2. Circuit diagram of a P-N-P heterostructure

In order to simplify this calculation, we have adopted several assumptions: (1) abrupt P-N junctions; (2) Depletion layer approximation; (3) Long-length approximation; (4) some assumptions for the bipolar transistor.

(1) Abrupt P-N junctions

We have divided this bipolar transistor into two coupled P-N junctions and assumed these P-N junctions are all abrupt junctions, which means that impurities are uniformly distributed in different regions. For one abrupt P-N junction, doping levels in P-type material and N-type material are kept constant when it is away from this junction and will abruptly change at the junction interfacial region (as in Figure 3b). It is a consensus that $N_a \gg N_d$ in the P-type region and $N_d \gg N_a$ in the N-type region. As a result physical parameters of the abrupt P-N junction usually vary in different regions, such as electric field intensity (as in Figure 3c)、 electric potential (as in Figure 3d) and doping levels.[22]

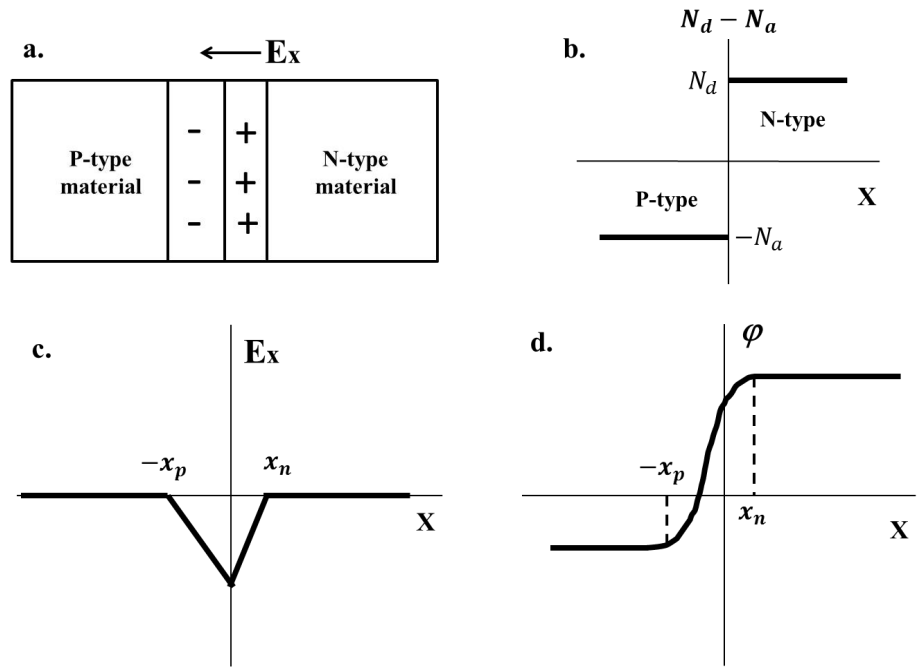

Figure 3. (a).An abrupt P-N junction; (b). Doping levels for the abrupt junction; (c). Electric field intensity for the abrupt junction; (d). Electric potential variation for the abrupt junction.

We have found that the electric field intensity is zero and the electric potential variation is constant when away from the junction interfacial region in thermal equilibrium. Therefore, we have divided this P-N junction into three regions: P-type region, N-type region and depletion layer, where the depletion layer has an electric potential difference and other regions are neutral.

(2) Depletion layer approximation

If P-type material and N-type material combine to form a P-N junction, the majority carriers of two regions will move to the opposite area due to carrier concentration gradients that exist at the interface. It results in a depletion layer at the interface when the equilibrium is reached again.

To solve for the potential barrier $V_{bi}$ across the depletion layer, we have employed the depletion layer approximation. In this assumption, there are no free electrons and holes in the depletion layer, only impurity ions ($N_d$ or $N_a$). Moreover, recombination is completely neglected in the depletion layer. As a result, the potential barrier $V_{bi}$ can be obtained by solving Possion's equation:[23]

$$\nabla^2 \varphi = -4\pi\rho / \varepsilon_r \varepsilon_0 \tag{1}$$

where $\rho = q\left(N_d - N_a\right)$ for the abrupt junction.

And the potential barrier $V_{bi}$ can be written as:

$$V_{bi} = \frac{k_B T}{q} \ln \frac{N_d N_a}{n_i^2} \tag{2}$$

We have also utilized another assumption that the voltage drop will only appear at the depletion layer if an external voltage ($V_s$) is applied. Simultaneously, the potential barrier is independent of the temperature gradient and external voltage.

(3) Long-length approximation

Generally, there are different Fermi levels between P-type material and N-type material. Therefore, the Fermi level can be realigned during the process of forming a PN junction, resulting in the quasi-Fermi level. To simplify this analysis, we have adopted the assumption that the quasi-Fermi level is constant.[24]

In reality, minority carriers of P-type material and N-type material need to be considered. This is because the recombination of minority carriers and majority carriers can affect the performance of a P-N junction. Therefore, the long-length approximation has been employed, in which the electron and hole diffusion lengths are shorter than the widths of P-type and N-type regions.[25] In other words, electrons and holes are recombined completely in the neutral region. The specific description is shown in Figure 4:

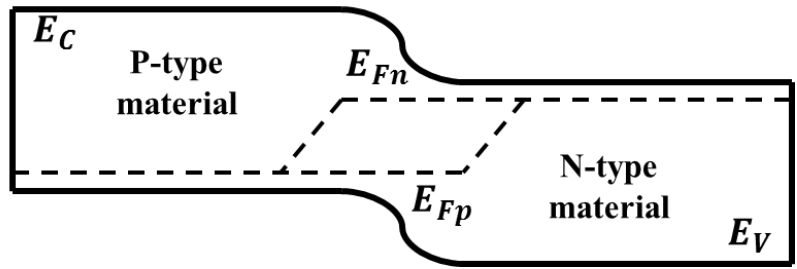

Figure 4: P-N junction in long-length approximation

The quasi-Fermi level ($E_{Fn}$) of N-type material is therefore constant in N-type region and the depletion layer under the same temperature, and the same applies to P-type material. The quasi-Fermi level is directly reflected in the dependence of the carrier concentration, resulting in a convenient calculation for the Seebeck coefficient:

$$S = \mp \frac{k_B}{q}\left[ \frac{E_F}{k_B T} - \left( r + \frac{5}{2} \right) \right] \tag{3a}$$

$$V_s = S\Delta T \tag{3b}$$

Where $r = -1/2$ is the phonon scattering parameter, $V_s$ is the Seebeck voltage, and $\Delta T$ is the temperature difference.

In this model, we have used Seebeck coefficients of bulk N-type material and P-type material as the basis for calculating the Seebeck voltage of the P-N junction.

(4 ) Some assumptions about the bipolar transistor

Although we have divided the bipolar transistor into two P-N junctions during the analysis process, the expression for volt-ampere characteristic of the bipolar transistor is still essential when calculating the overall thermoelectric performance. The suitable expression for volt-ampere characteristic of the bipolar transistor is not only based on above assumptions, but also requires some further hypotheses.

In our model, the base width is very thin(around 20nm) and the recombination of electrons and holes in the base region is neglected so that the base current is equal to zero, which means that the emitter current is equal to the collector current. Meanwhile, we have adopted small injection levels. Therefore, the expression for volt-ampere characteristic of the bipolar transistor is written as:

$$J_E = -q \left( \frac{D_{pB} P_{nB}^\circ}{W_B} + \frac{D_{nE} N_{pE}^\circ}{L_{nE}} \right) \left( e^{qV_E / k_B T} - 1 \right) \tag{4}$$

where $J_E$ is the emitter current density, $D_{pB} = \frac{k_B T}{q} \mu_{pB}$ is the hole diffusion coefficient of the base material, $\mu_{pB}$ is the minority carrier mobility of the base material, $P_{nB}^\circ$ is the minority carrier concentration of the base material, $W_B$ is the base width, $D_{nE} = \frac{k_B T}{q} \mu_{nE}$ is the electron diffusion coefficient of the emitter material, $\mu_{nE}$ is the minority carrier mobility of the emitter material, $N_{pE}^\circ$ is the minority carrier concentration of the emitter material, $L_{nE} = \sqrt{D_{nE} \tau_{nE}}$ is the minority carrier diffusion length of the emitter material, $\tau_{nE}$ is the relaxation time of the emitter material, and $V_E$ is the emitter voltage.

We have assumed the emitter current is equal to the collector current, and the current density of both is the same if junction areas of the base-emitter interface and base-collector interface are identical. Therefore, $|J_E| = |J_C|$ can be obtained.

Subsequently, the collector voltage and emitter voltage can thus be determined. We have assumed that the emitter is forward biased and the collector is reverse biased. In addition, the Seebeck voltage caused by the temperature gradient is treated as an external power supply for the bipolar transistor. Moreover, the potential barrier $V_{bi}^E$ of the base-emitter interface and the potential barrier $V_{bi}^C$ of the base-collector interface are independent of the external voltage. Therefore, the emitter voltage $V_E$ and collector voltage $V_C$ can be written as:

$$V_E = V_{bi}^E - S_E \Delta T \tag{5a}$$

$$V_C = V_{bi}^C + S_C \Delta T \tag{5b}$$

where $S_E$ is the Seebeck coefficient of the emitter material, and $S_C$ is the Seebeck coefficient of the collector material.

After determining $V_E$ and $V_C$, the depletion layer width ( $x_{dC}$ ) across the base-collector interface can be written as:

$$x_{nC} = \sqrt{\frac{2\varepsilon_r \varepsilon_0 V_C N_a}{q N_d \left( N_a + N_d \right)}} \tag{6a}$$

$$x_{pC} = \sqrt{\frac{2\varepsilon_r\varepsilon_0 V_C N_d}{qN_a\left(N_a + N_d\right)}} \qquad (6b)$$

$$x_{dC} = x_{nC} + x_{pC} \qquad (6c)$$

where $x_{nC}$ is the depletion layer width near the base region, and $x_{pC}$ is the depletion layer width near the collector region. And the depletion layer width across the base-emitter interface ( $x_{dE} = x_{nE} + x_{pE}$ ) can also be obtained. Consequently, the base width in Eq.(4) is equal to $x_{nE} + x_{nC}$.

The electrical conductivity of the bipolar transistor can be derived from the emitter current density:

$$\sigma = \frac{|J_e|}{\overline{E}} \qquad (7)$$

where $\overline{E} = \dfrac{V_E}{x_{nE} + x_{pE}}$ is the average electric field intensity across the base-emitter interface (depletion layer) as the electric field varies linearly with distance for an abrupt P-N junction.

The Seebeck coefficient of the bipolar transistor can be obtained based on the collector voltage:

$$S_{PNP} = \frac{V_c}{\Delta T} \qquad (8)$$

Finally, the power factor and ZT of the bipolar transistor can also be obtained:

$$P = \sigma S_{PNP}^2 \qquad (9a)$$

$$ZT = \frac{\sigma S_{PNP}^2}{\lambda} T \qquad (9b)$$

By solving Eq.(3a)-(9b), the thermoelectric performance of the bipolar transistor can be obtained.

## III. Result and Discussion

The model described in this work has been implemented using MATLAB. Because $Bi_2Te_3$ and $Sb_2Te_3$ are widely used thermoelectric semiconductor materials, we applied our model to a P-$Bi_{0.5}Sb_{1.5}Te_3$/N-$Bi_2Te_3$/ P-$Bi_{0.5}Sb_{1.5}Te_3$ heterostructure in order to evaluate the potentiality of this model. Table II. shows these material parameters used in the simulation.

Table II. Material parameters used for the simulation ($T_c$=298K $T_h$=323.15K)

| Material | Emitter | Base | Collector |
|---|---|---|---|
| Parameter | (P-$Bi_{0.5}Sb_{1.5}Te_3$) | (N-$Bi_2Te_3$) | (P-$Bi_{0.5}Sb_{1.5}Te_3$) |
| Eg(eV) | 0.18[26] | 0.17[27] | 0.18 |
| $\varepsilon_r$ | 260[26] | 400[28] | 260 |
| m*/m0 | 1.441[29] | 1.4[30] | 1.467[29] |
| $\mu_n$(cm$^2$/V s) | 120[31] | 1200[17] | 120 |
| | 8000[32] | \ | 8000[32] |
| $\mu_p$(cm$^2$/V s) | \ | 510[17] | \ |
| | \ | 4778.3[33] | \ |
| $D_n$(cm$^2$/s) | 3.3446 | 33.466 | 3.3446 |
| | 222.97 | \ | 222.97 |
| $D_p$(cm$^2$/s) | \ | 14.2146 | \ |
| | \ | 133.18 | \ |
| $\tau_{nE}$($\mu$s) | 0.24[32] | \ | 0.24 |
| $L_{nE}$($\mu$m) | 12.7 | \ | 12.7 |
| | 103.5 | \ | 103.5 |
| $n_i$(cm$^{-3}$) | 1.5×10$^{18}$ | 1.72×10$^{18}$ | 1.5×10$^{18}$ |

Figure 5 shows the relationship between Seebeck coefficients and doping concentrations for the emitter material, base material and collector material according to Eq.(3a). In order to obtain the better thermoelectric performance, we set the range of Seebeck coefficients for these materials at ±220μV/K～±240μV/K. Therefore, the scope of doping concentrations for these materials can subsequently be determined. These results are derived from experimental data found in previous literature.[29,30]

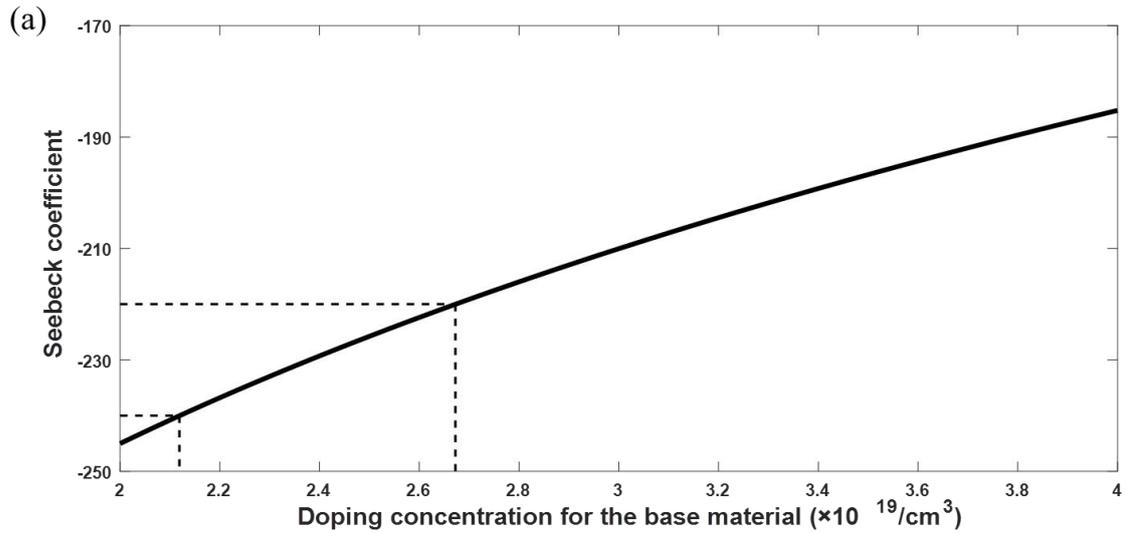

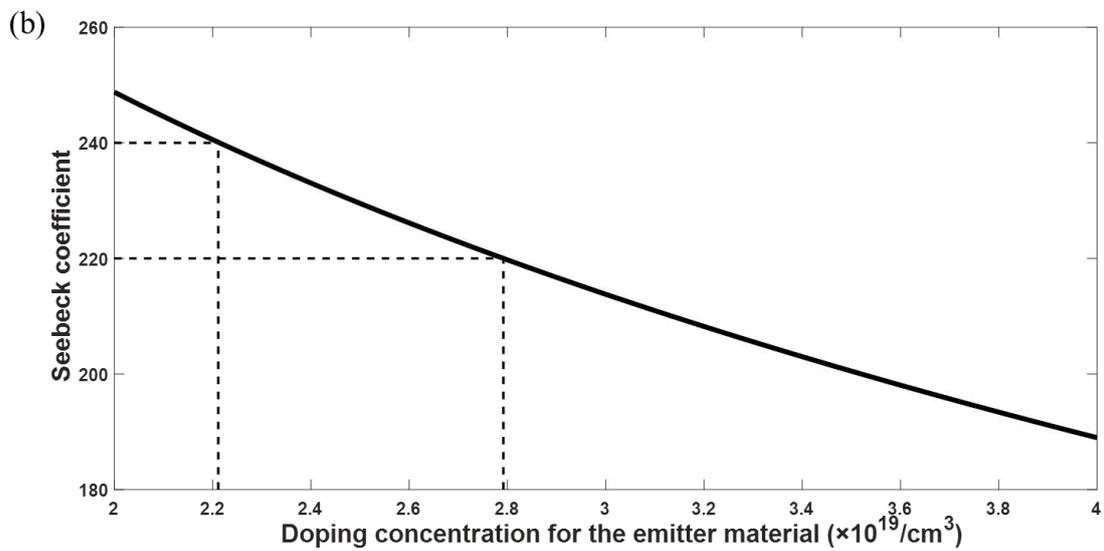

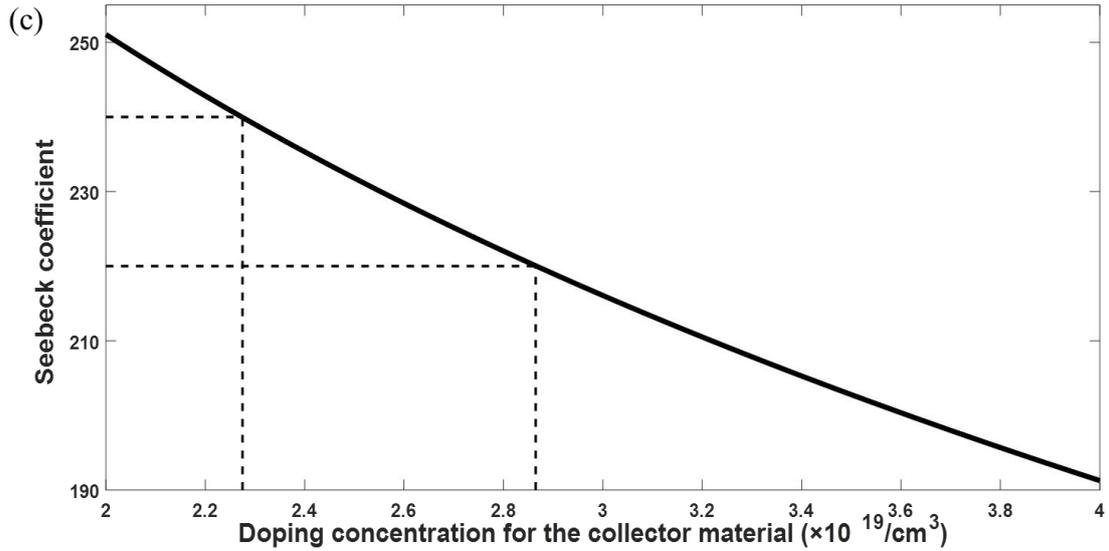

Figure 5. The relationship between Seebeck coefficients and doping concentrations for (a) the base material, (b) the emitter material, and (c) the collector material

We could estimate the range of doping concentrations for these materials through calculated results. The range for base material, emitter material, and collector material is $2.1 \times 10^{19} \sim 2.7 \times 10^{19} / cm^3$, $2.2 \times 10^{19} \sim 2.8 \times 10^{19} / cm^3$ and $2.2 \times 10^{19} \sim 2.8 \times 10^{19} / cm^3$ respectively.

## 2.1 Collector voltage

According to Eq. (5b), the collector voltage is the sum of the potential barrier of the base-collector interface and the Seebeck voltage. Moreover, we considered that the collector voltage is a function of the base doping concentration and collector doping concentration based on Eq.(2) and Eq.(3a), as shown in Figure 6.

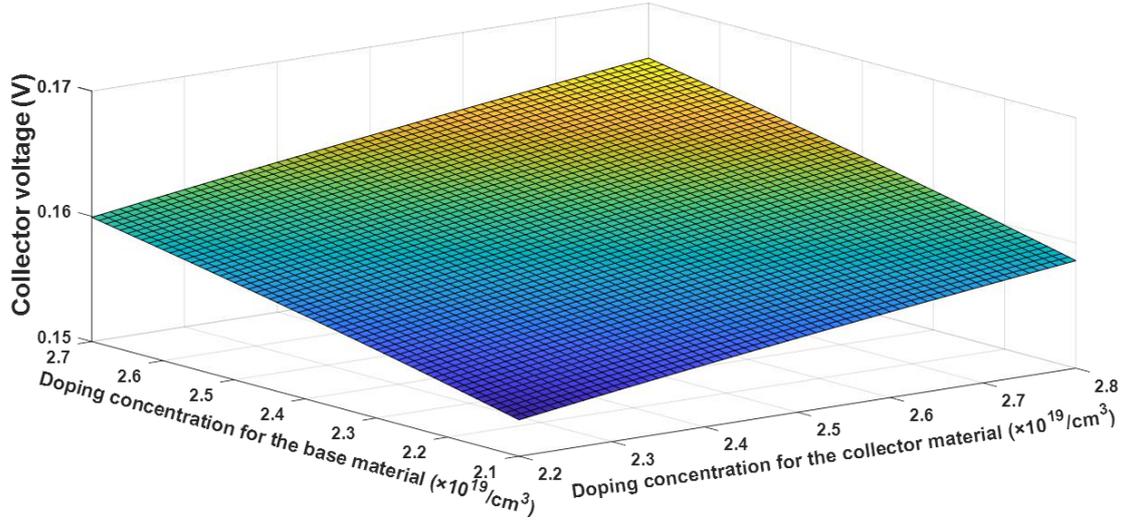

Figure 6. Collector voltage in the P-N-P heterostructure

In order to obtain the maximum output power, the collector voltage should be at the maximum. We can see that the maximum collector voltage is 0.1656V when doping concentrations of the base material and collector material are $2.7 \times 10^{19} / cm^3$ and $2.8 \times 10^{19} / cm^3$, respectively. Therefore, we can determine the optimal doping concentrations for the base material and collector material. Subsequently, $x_{nC} = 11.75$ nm can be obtained to calculate the base width.

## 2.2. Emitter voltage and Emitter current density

After determining the base doping concentration, the emitter voltage is also derived from Eq.(2), (3a) and (5a). Figure 7 shows the relationship of the emitter voltage and the emitter doping concentration.

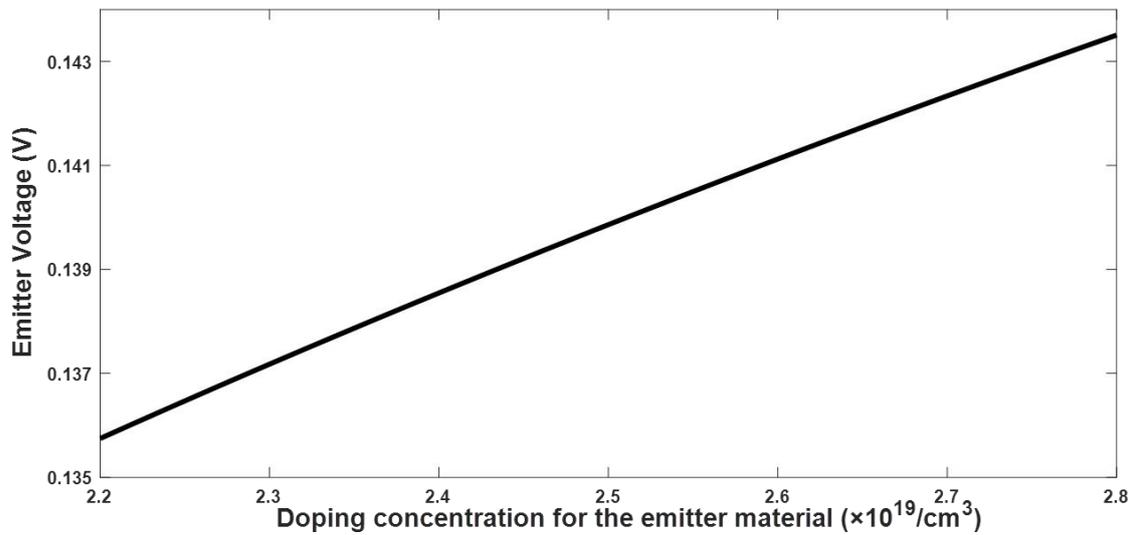

Figure 7. Emitter voltage in the P-N-P heterostructure

We find that the magnitude of the emitter voltage increases as the emitter doping concentration increases. This finding indicates that the potential barrier of the base-emitter interface is positively correlated with the emitter doping concentration and the Seebeck voltage is relatively small. But we can't define the optimal doping concentration of the emitter material just from Figure 8 as the emitter current density is a more important parameter for the bipolar transistor. As a result, we calculated the emitter current density based on data presented in Table II. The relationship between the emitter current density and the emitter doping concentration under different carrier mobility is shown in Figure 8.

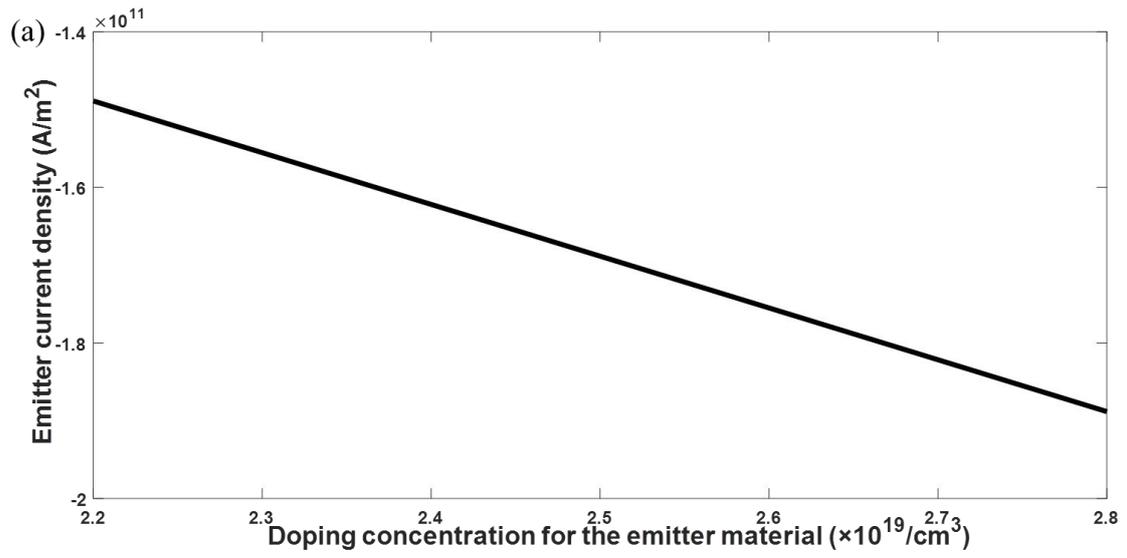

(a)

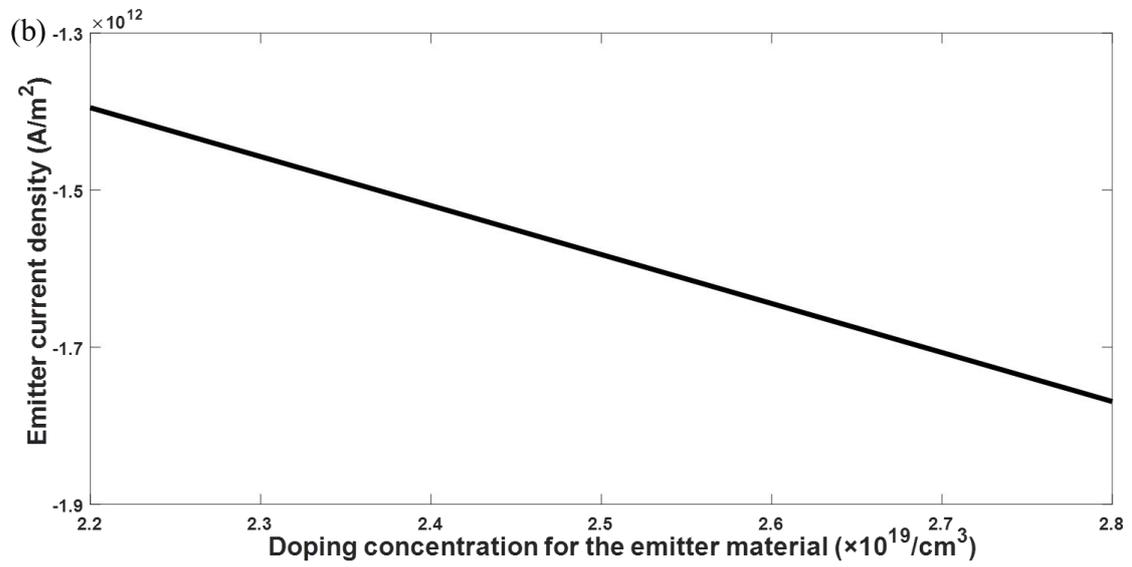

(b)

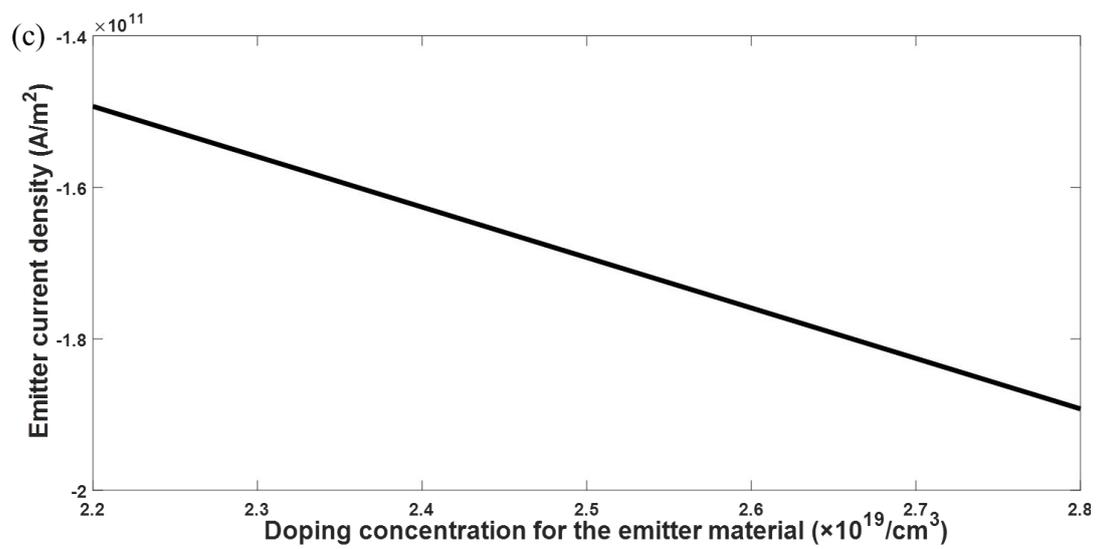

(c)

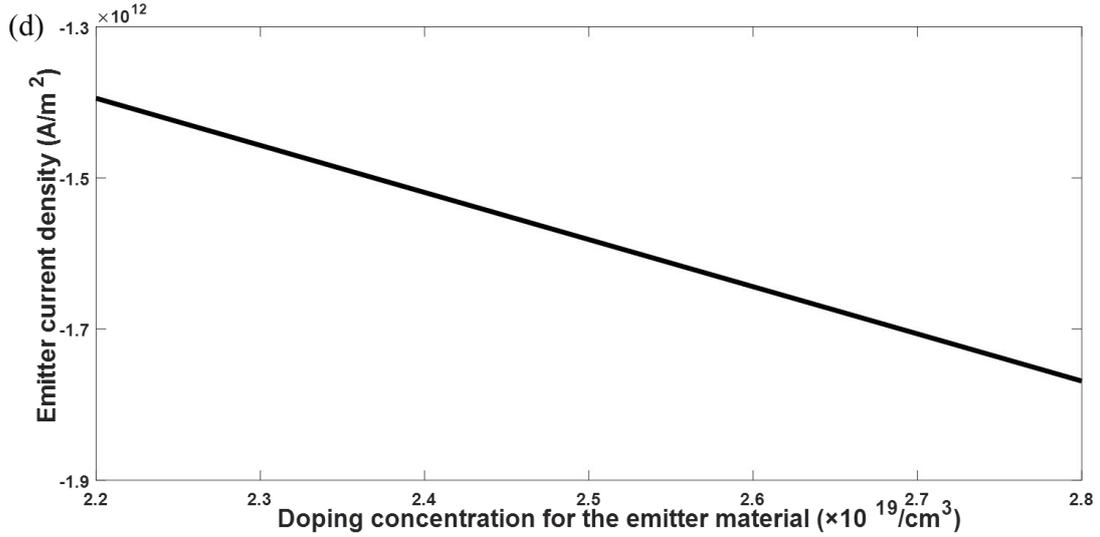

Figure 8. Emitter current density in the P-N-P heterostructure

(a) $\mu_p = 510 cm^2 \cdot V^{-1} \cdot s^{-1}$ for the base material, $\mu_n = 120 cm^2 \cdot V^{-1} \cdot s^{-1}$ for the emitter material (b) $\mu_p = 4778 cm^2 \cdot V^{-1} \cdot s^{-1}$ for the base material, $\mu_n = 8000 cm^2 \cdot V^{-1} \cdot s^{-1}$ for the emitter material (c) $\mu_p = 510 cm^2 \cdot V^{-1} \cdot s^{-1}$ for the base material, $\mu_n = 8000 cm^2 \cdot V^{-1} \cdot s^{-1}$ for the emitter material (d) $\mu_p = 4778 cm^2 \cdot V^{-1} \cdot s^{-1}$ for the base material, $\mu_n = 120 cm^2 \cdot V^{-1} \cdot s^{-1}$ for the emitter material

As it can be seen, there are different emitter current densities under distinct carrier mobility and the absolute values of emitter current densities increase with the increase of the emitter doping concentration. In order to gain the maximum output power, we determined the emitter current density should be the maximum and also obtained the optimal emitter doping concentration along with the maximum emitter voltage. Meanwhile, the depletion layer width ( $x_{nE} + x_{pE}$ ) across the base-emitter interface and the base width ($x_{nE} + x_{nC}$) were determined.

We summarized the maximum emitter current densities under different conditions in Table III.

Table III. The maximum emitter current densities under different carrier mobility (T=323.15K)

| Carrier Mobility | EmitterCurrent Density/A/m$^2$ | Doping Concentration/cm$^3$ | Emitter Voltage/V | Base Width/nm |
|---|---|---|---|---|
| $\mu_p$=510cm$^2$/V s $\mu_n$=120cm$^2$/V s | -1.888×10$^{11}$ | 2.8×10$^{19}$ | 0.1435 | 22.69 |
| $\mu_p$=4778cm$^2$/V s $\mu_n$=8000cm$^2$/V s | -1.769×10$^{12}$ | 2.8×10$^{19}$ | 0.1435 | 22.69 |
| $\mu_p$=510cm$^2$/V s $\mu_n$=8000cm$^2$/V s | -1.892×10$^{11}$ | 2.8×10$^{19}$ | 0.1435 | 22.69 |
| $\mu_p$=4778cm$^2$/V s $\mu_n$=120cm$^2$/V s | -1.769×10$^{12}$ | 2.8×10$^{19}$ | 0.1435 | 22.69 |

It is evident that the maximum emitter current densities remain almost unchanged when the minority carrier mobility ($\mu_p$) of the base material is constant from Table III. This is because the base width is nanoscale and the electron diffusion length of the emitter material is micron level. Therefore, we considered that the minority carrier mobility of the base material is the main factor affecting the maximum emitter current densities. Moreover, the nanoscale base width results in a large emitter current density, which increases the output power in the bipolar transistor.

## 2.3 Thermoelectric performance

Because we selected the optimal collector voltage and the maximum emitter current density, the optimal electrical conductivity and Seebeck coefficient of the bipolar transistor can be obtained according to Eq.(7) -(9b). Specific calculation results are shown in Table IV.

Table IV. Electrical conductivities, Seebeck coefficients, power factors and ZT for the P-N-P heterostructure (T=323.15K)

| Case | Carrier Mobility /cm²/V s | Electrical Conductivity /S/cm | Seebeck Coefficient /μV/K | Power Factor /μW/cm K² | $ZT_{max}$ |
|------|---------------------------|-------------------------------|---------------------------|------------------------|------------|
| I | $\mu_p$=510 $\mu_n$=120 | 255.9 | 3312 | 2807.06 | 45 |
| II | $\mu_p$=4778 $\mu_n$=8000 | 2397.7 | 3312 | 26301.2 | 425 |
| III | $\mu_p$=510 $\mu_n$=8000 | 256.44 | 3312 | 2813 | 45 |
| IV | $\mu_p$=4778 $\mu_n$=120 | 2397.7 | 3312 | 26301.2 | 425 |

The electrical conductivities of all the cases are shown in Table IV. We can find that these cases have electrical conductivities of about 255.9S/cm, 2397.7S/cm, 256.44S/cm and 2397.7S/cm, respectively. The magnitude of electrical conductivity of all the cases change with the hole mobility of base material (N-type $Bi_2Te_3$). Compared with N-type $Bi_2Te_3$ bulk material ($\sigma=1204.5 S/cm$)[29] and film ($\sigma=2174 S/cm$)[34] in previous literature, electrical conductivities of Case I and Case III are lower than those of the bulk material and film. Because electrical conductivities are derived from Eq.(4) and Eq.(7), we consider that the emitter current density is the main factor affecting electrical conductivities. We can see that the emitter current density is related to minority carrier concentrations at the base region and the emitter region in Eq.(4). Therefore, possible explanation behind the low electrical conductivities of Case I and Case III is the relatively low minority carrier concentrations at the base-emitter interface. On the other hand, Case II and Case IV have almost the same electrical conductivities as films. This is because the minority carrier mobility of these two cases is larger than those of films ( $\mu = 10.7 cm^2 / Vs$ ) in Reference [34]. We believe that the reduction in the carrier concentration is more than compensated for by the increase in the carrier mobility.

The Seebeck coefficients of these cases exhibit very large absolute values, which are 50 times greater than those of films and 15 times greater than those of bulk materials. We believe that such a large Seebeck coefficient may be caused by the potential barrier of the depletion layer in the base-collector interface. In this depletion region, majority carriers can be depleted completely and the recombination of electrons in N-type material and holes in P-type material at the cold region under this temperature gradient can be eliminated. Therefore, the high values of Seebeck coefficients can be obtained.

So far, we have considered only electrical conductivities and Seebeck coefficient since it is through these performances that our model predicts an increase in ZT. However, ZT is also in dependence of the thermal conductivity κ. If we assume that the thermal conductivity of the bipolar transistor is equal to 2W/m K, our calculated results in Table III imply that the bipolar transistor ZT can be up to dozens of times or even hundreds of times greater than bulk and film values, giving values of $ZT_{max} = 45$ in Cases I and III and $ZT_{max} = 425$ in Cases II and IV at 323.15K.

## IV. Summary

In this paper, we present a general model for P-N-P heterostructures. In our model, P-N-P heterostructures behave as bipolar transistors due to the setting of temperature gradient. We take an abrupt P-Bi$_{0.5}$Sb$_{1.5}$Te$_3$/N-Bi$_2$Te$_3$/ P-Bi$_{0.5}$Sb$_{1.5}$Te$_3$ heterostructure as example to obtain thermoelectric performance of this model. The electrical conductivity based on the emitter current density and the Seebeck coefficient based on the collector voltage for the bipolar transistor were calculated in order to control the Seebeck coefficient and the electrical conductivity independently. The conclusion can be obtained as below.

(1) With the optimal doping concentration of $2.7 \times 10^{19} / cm^3$, $2.8 \times 10^{19} / cm^3$ and $2.8 \times 10^{19} / cm^3$ for the base material, emitter material and collector material, respectively, the maximum emitter current density and the maximum collector voltage under different carrier mobility are 1.892 $\times 10^{11}$A/m$^2$ or 1.769 $\times 10^{12}$A/m$^2$ and 0.1656V. The $ZT_{max}$ values can reach 45 or 425 at 323.15K. Therefore, this model is a useful method to enhance thermoelectric performances and efficiency.

(2) This P-N-P heterostructure model not only provides a completely new perspective for research in the thermoelectric field, but also for other energy conversion mechanisms, such as photovoltaic conversion.

Acknowledgements:

This work was financially supported by The National Key Research

and Development Program of China [Grant No. 2017YFF0204706], by the Fundamental Research Funds for the Central Universities [Grant No. FRF-MP-18-005, and FRF-MP-19-005].

## Appendix: Emitter current density for an idealized silicon N-P-N bipolar transistor

The emitter current density is written as:

$$J_E = -q\left(\frac{D_{nB}n_{pB}^{\circ}}{W_B} + \frac{D_{pE}n_{nE}^{\circ}}{L_{pE}}\right)\left(e^{qV_E/k_BT} - 1\right) \tag{A1}$$

The relevant data are as follow:[22]

$n_i = 2.26 \times 10^{11}/cm^3$ , $n_{pb}^{\circ} = \dfrac{n_i^2}{N_{aB}} = \dfrac{\left(2.26 \times 10^{11}\right)^2}{5 \times 10^{16}} = 1.02 \times 10^6/cm^3$ , $W_B = 500nm$ ,

$\tau_n = 10^{-6}s$ in the base , $\tau_p = 5 \times 10^{-7}s$ in the emitter, $D_{nB} = 22.5cm^2 \cdot s^{-1}$ ,

$D_{pE} = 6.5cm^2 \cdot s^{-1}$ , $L_p = 18\mu m$ ,

$n_{ne}^{\circ} = \dfrac{n_i^2}{N_{dE}} = \dfrac{\left(2.26 \times 10^{11}\right)^2}{2 \times 10^{17}} = 2.55 \times 10^5/cm^3$ , $J_E = -1.44 \times 10^8 A/m^2$ is shown in the literature.

We utilize Eq.(A1) to obtain the emitter current density $J_E^{'}$ based on these data and compare with $J_E$ .

Therefore, $J_E^{'} = -1.41 \times 10^8 A/m^2$ is obtained. And the result is very close to $J_E$ .

Finally, we can conclude that Eq.(4) and Eq.(A1) are all effective for calculating emitter current densities of P-N-P bipolar transistors and N-P-N bipolar transistors.

2013.